\documentclass[useAMS,usenatbib]{mn2e}
\usepackage{mathrsfs}
\usepackage{graphicx,amssym,color}
\newcommand{\beq}{\begin{equation}}
\newcommand{\eeq}{\end{equation}}
\newcommand{\beqa}{\begin{eqnarray}}
\newcommand{\eeqa}{\end{eqnarray}}

\usepackage{color}
\def\gsim { \lower .75ex \hbox{$\sim$} \llap{\raise .27ex \hbox{$>$}} }
\def\lsim { \lower .75ex \hbox{$\sim$} \llap{\raise .27ex \hbox{$<$}} }
\def\proptosim { \lower .75ex \hbox{$\sim$} \llap{\raise .27ex \hbox{$\propto$}} }

\newcommand{\bc}{\begin{center}}
\newcommand{\ec}{\end{center}}

\title[Galaxy Pairs in the Local Group]
      {Galaxy Pairs in the Local Group}
\author[Fattahi et al. ]{
\parbox[t]{\textwidth}{
       Azadeh Fattahi$^{1}$\thanks{Email: azadehf@uvic.ca}, Julio
         F. Navarro$^1$, Else Starkenburg$^1$, Christopher
         R. Barber$^1$, and Alan
         W. McConnachie$^2$
}
  \\ \\
$^1$Department of Physics and Astronomy, University of Victoria, PO
Box 3055 STN CSC, Victoria, BC, V8W 3P6, Canada\\
$^2$NRC Herzberg Institute of Astrophysics, 5071 West Saanich Road, Victoria, BC, V9E 2E7, Canada\\
}

\begin{document}

\date{\today}

\pagerange{\pageref{firstpage}--\pageref{lastpage}}
\pubyear{2013}

\maketitle

\label{firstpage}

\begin{abstract}
  Current models of galaxy formation predict that galaxy pairs of
  comparable magnitudes should become increasingly rare with
  decreasing luminosity. This seems at odds with the relatively high
  frequency of pairings among dwarf galaxies in the Local Group. We
  use literature data to show that $\sim 30\%$ of all satellites of
  the Milky Way and Andromeda galaxies brighter than $M_{V}=-8$ are
  found in likely physical pairs of comparable luminosity. Besides the
  previously recognised pairings of the Magellanic Clouds and of NGC 147/NGC 185,
  other candidate pairs include the Ursa Minor and Draco dwarf
  spheroidals, as well as the And I/And III satellites of M31. These
  pairs are much closer than expected by chance if the radial and
  angular distributions of satellites were uncorrelated; in addition,
  they have very similar line-of-sight velocities and luminosities that differ by
  less than three magnitudes. In contrast, the same criteria pair
  fewer than $4\%$ of satellites in N-body/semi-analytic models that
  match the radial distribution and luminosity function of Local Group
  satellites. If confirmed in studies of larger samples, the high
  frequency of dwarf galaxy pairings may provide interesting clues to
  the formation of faint galaxies in the current cosmological
  paradigm.
\end{abstract}


\section{Introduction}
\label{SecIntro}


The shapes of the galaxy and dark halo mass functions differ
substantially in the $\Lambda$CDM paradigm \citep[see,
e.g.,][]{Benson2003}. This is usually interpreted to imply that the
``efficiency'' of galaxy formation, as measured by the ratio between
the stellar mass of a galaxy ($M_{\rm gal}$) and the
virial\footnote{We define all virial quantities as those corresponding
  to a sphere of mean density equal to $200$ times the critical
  density for closure. Furthermore, we shall hereafter refer to the
  stellar mass of a galaxy as ``galaxy mass'', for brevity.}  mass of
its host halo ($M_{200}$), varies strongly with virial mass. In
particular, $M_{\rm gal}/M_{200}$ should decrease steeply toward low
halo masses in order to match the shallow faint end of the galaxy
luminosity function \citep[see,
e.g.,][]{Moster2010,Behroozi2010,Guo2011}.

On the scale of dwarf galaxies, which we define conventionally here as
those with $M_{\rm gal}<10^{9.5}\, M_\odot$, simple
abundance-matching models suggest a dependence nearly as steep as
$M_{\rm gal} \propto M_{200}^{3}$ in the dwarf galaxy regime
\citep{Guo2010}. Such steep scaling would imply that dwarfs spanning
several decades in stellar mass should nevertheless inhabit halos of
similar virial mass. In addition, extrapolating such models to the faintest
galaxies known indicate that few, if any, galaxies more massive
than a few million solar masses are expected to form in halos with virial mass below
$10^{10}\, M_\odot$.

Recent work has highlighted potential disagreements between these
model predictions and observations, including the lack of a
characteristic velocity at the faint-end of blind HI surveys 
\citep{Zwaan2010}; and the low virial mass
(substantially below $10^{10}\, M_\odot$) inferred from dynamical data
for the dwarf spheroidal companions of the Milky Way
\citep{Boylan-Kolchin2012} and for nearby dwarf irregulars
\citep{Ferrero2012}. It could be argued, however, that the evidence
for substantial disagreement is unconvincing, given that the
inferences are either indirect (in the case of the HI velocity
function) or based on small and heterogeneous samples (in the case of
the nearby dwarfs; see \citealt{Wang2012a,Vera-Ciro2012}).

It is therefore important to consider further tests of the model
predictions.  We explore here how the steep $M_{\rm gal}$-$M_{200}$
relation predicted for dwarfs affects the frequency of galaxy pairs of
comparable luminosity. Such pairs, when close enough to inhabit the
same dark matter halo (referred to hereafter as ``physical pairs''),
are expected to be rare at all luminosities, but especially so in the
scale of dwarfs. This is because the
fainter companions in physical pairs trace the halo substructure, and
subhalos are, by and large, far less massive than the main
halo: the most massive subhalo typically has a mass only one
hundredth that of the main system \citep[see,
e.g.,][]{Springel2008b,Wang2012a}.

Pairs of comparable luminosity are therefore more likely to form in
fairly massive halos, where galaxy formation efficiency decreases
with increasing halo mass, partly compensating the mass difference
between the main halo and its most massive subhalo. On dwarf galaxy
scales the situation is reversed, and the precipitous decline in
galaxy formation efficiency with decreasing halo mass should curb the
formation of physical pairs of comparable luminosity. More generally
speaking, isolated associations of dwarf galaxies should be rare. They
are known to exist \citep[e.g.,][]{Tully2006,Soares2007}, but their cosmological
abundance and dependence on luminosity have not yet been adequately
established \citep[see][for a recent attempt]{Sales2012}.

The Local Group offers an interesting environment to test these
ideas. Advantages include the fact that, away from the ``zone of
avoidance'' caused by Galactic dust, the census of Milky Way (MW)
satellites brighter than $M_V\sim -8$ is complete \citep[see,
e.g.,][]{Whiting2007}, and that accurate magnitudes, positions,
distances, and line-of-sight velocities are known for all. Many
fainter systems in the Local Group still remain undiscovered, as demonstrated by recent
discoveries both by the Sloan Digital Sky Survey in the Milky Way
\citep[][]{Koposov2008}, and by the Pan-Andromeda
Archaeological Survey around M31
\citep[PAndAS;][]{McConnachie2009}. In the latter the census is likely
complete within the PAndAS survey area down to a magnitude limit of
$M_V \sim -6.5$, although some brighter dwarfs at larger radius likely
have yet to be identified \citep[as exemplified by the discoveries of
Andromeda XXVIII and XXIX;][]{Bell2011,Slater2011}.

Further, we know that at least some of the satellites of the Milky Way
and M31 are very likely physically associated and bound to each
other. An obvious pairing is that of the Magellanic Clouds \citep[see,
e.g.,][and references therein]{Kallivayalil2006b}. Around M31, there
have been suggestions that NGC 147 and NGC 185 also form a bound pair
\citep{vandenBergh1998}. If these pairs are bound, their vicinity to
their primary galaxy suggests that we are observing them just before
they are separated by the tidal field of the main galaxy
\citep{Besla2007,Sales2011}. This implies very recent accretion and
indicates that their occurrence should not be uncommon amongst
isolated systems.

Taken at face value, the existence of these two pairs of dwarfs seems
at odds with the expected rarity of such associations. We use this as
motivation to search, using literature data, for other dwarf galaxy
pairs in the Local Group. We describe in Sec.~\ref{SecData} the observational
 dataset and the simulated satellite dataset we use for comparison.
 In Sec.~\ref{SecRes} we introduce the pairing procedure we have adopted 
and compare the results with those obtained when the same procedure 
is applied to a hybrid N-body/semi-analytic model of satellite galaxy 
formation applied to N-body simulations from the Aquarius Project. 
We conclude with a brief summary in Sec.~\ref{SecConc}.

\section{Datasets}
\label{SecData}

We use the recent compilation by \citet{McConnachie2012} as the source
of the positions, distances, line-of-sight velocities, and magnitudes
of Local Group dwarfs that we use in our analysis. All velocities are
heliocentric and corrected to the rest frame of the Galaxy. In order
to prevent biases due to incompleteness, we consider only satellites
brighter than $M_V=-8$ located within $300$ kpc of the Milky Way or
Andromeda (M31) galaxies. The sample consists of $29$ dwarfs, $17$ of
which orbit around M31; the rest are satellites of the Milky Way.  

For comparison, we have identified analogous samples of simulated
satellites in the six $\sim 10^{12}\, M_\odot$ halos of the Aquarius
Project \citep{Springel2008b} using the model of
\citet{Starkenburg2012}. This is a semi-analytic model grafted onto
the level-2 Aquarius runs, which simulate each halo with several
hundred million particles, thus ensuring a resolution high enough to
track the formation of all halos and subhalos that might plausibly
host the dwarf galaxies brighter than $M_V=-8$ we use in our
analysis. The model satellites of each Aquarius halo have luminosity
and radial distributions that are broadly consistent with the Milky
Way and M31 and therefore provide a useful testbed of the significance
of our results for $\Lambda$CDM dwarf galaxy formation models.

There are a total of $175$ simulated satellites brighter than $M_V=-8$
within $300$ kpc of the primary galaxies of all six Aquarius 
halos (on average $29$ per halo). The model provides not only the 
full 3D position and
velocity information for all of them, but also allows us to track
their evolution. Our simulated sample does not include satellites
whose dark matter halos have been fully disrupted by tides, since
their fate is uncertain.  We refer the interested reader to
\citet{Starkenburg2012} for details on the semi-analytic model.

\begin{figure}
  \bc \hspace{-0.2cm}
  \resizebox{8.5cm}{!}{\includegraphics{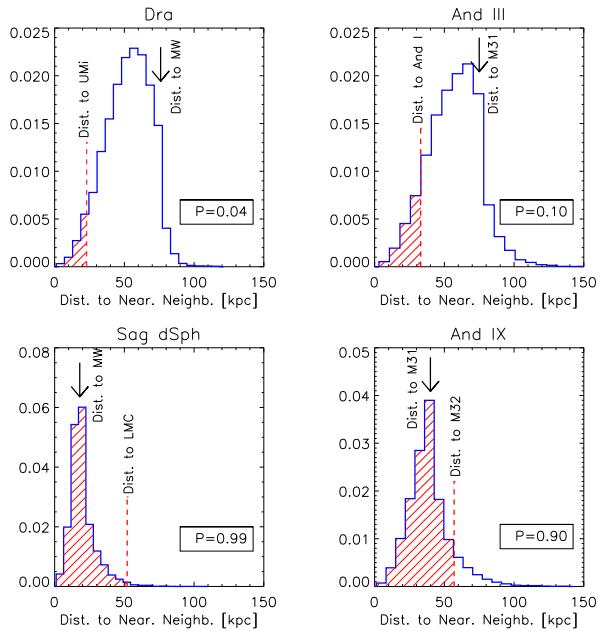}}\\%
  \caption{ Distribution of nearest-neighbour distances, $d_{\rm nn}$,
    to two Milky Way satellites (panels on the left) and two M31
    satellites (panels on the right) expected if satellites were
    distributed isotropically about each primary with a radial
    distribution consistent with the observed one. The probability, $P$,
    that a satellite's nearest neighbour lies, by chance, as close as
    or closer than observed is highlighted by the shaded region of
    each histogram and quoted in each panel's legend. A downward arrow
    indicates the distance to the primary galaxy. The top panels
    illustrate cases where the probability is rather small, indicative
    of a potential physical association. The bottom panels, on the
    other hand, illustrate two cases where the nearest neighbours
    are not significantly closer than expected at random. }
\label{FigDnn} \ec
\end{figure}

\begin{figure}
\bc
\hspace{-0.5cm}
\resizebox{8.5cm}{!}{\includegraphics{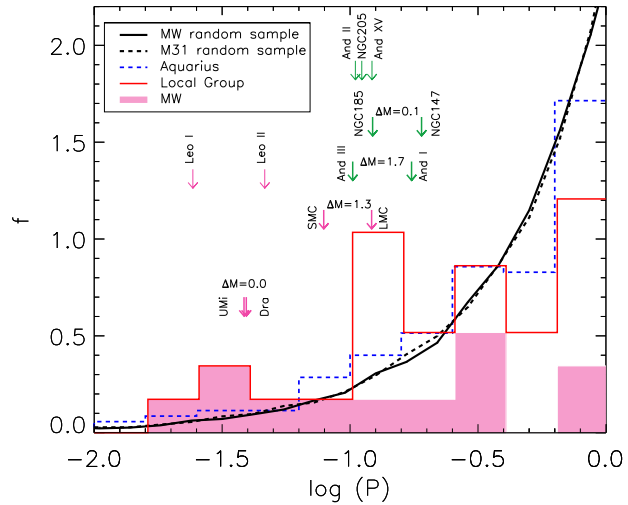}}\\%
\caption{Distribution of the probability, $P$, of having a nearest
  neighbour as close as or closer than observed if the satellites were
  isotropically distributed around each primary and had the same
  radial distribution as that of the Milky Way (solid thick line) and
  M31 (solid dashed line). Only $20\%$ of satellites are expected to
  have $P<0.2$, with a very weak dependence on the number of
  satellites and the shape of the radial profile. The probability
  distribution obtained for a semi-analytic model applied to the six
  Galaxy-sized halos of the Aquarius Project is shown by the dashed
  blue histogram; $48$ out of $175$ satellites have $P<0.2$, or $27\%$
  of the total. The corresponding distribution for Local Group
  satellites is shown by the solid red histogram (the contribution of
  Milky Way satellites is highlighted by the shaded area of the
  histogram). In the Local Group, more than $40\%$ of satellites have
  $P<0.2$, a result expected to happen by chance in fewer than one in
  $100$ random realizations. The angular and radial distributions of
  satellites thus seem highly correlated and suggest the presence
  of physically-associated pairs.  }
\label{FigP}
\ec
\end{figure}

\section{Analysis and Results}
\label{SecRes}


It has long been noticed that the spatial distribution of Milky Way
satellites is highly anisotropic \citep{Lynden-Bell1976}, and is often
described as a polar plane whose significance has been the matter of
much recent debate \citep[see,
e.g.,][]{Kroupa2005,Zentner2005,Libeskind2005,Metz2007}. Around M31,
14 out of 17 satellites in our sample are in the hemisphere nearer the
Milky Way \citep{McConnachie2006}. Further, several have recently been 
shown to delineate a flattened structure that in total comprises at
least half of all known M31 satellites \citep{Ibata2013}.
These are unlikely configurations for a virialized population and hint
strongly at recent accretion.

Our pairing procedure begins by identifying satellites whose nearest
neighbour is unusually close when compared with the probability
distribution of nearest-neighbour distances, $d_{\rm nn}$, obtained by
Monte Carlo sampling a random isotropic population of satellites with
the same total number and radial distribution. We illustrate this in
Fig.~\ref{FigDnn}, where we show the $d_{\rm nn}$ distribution
expected for two satellites of the Milky Way (Draco and Sagittarius),
and two of M31 (And III and And IX). 

The bottom left panel of Fig.~\ref{FigDnn} shows that the nearest
satellite to Sagittarius (the LMC, $52$ kpc away) is about twice as
far as the distance at which the probability distribution of
nearest-neighbour distances peaks.  If the radial and angular
distribution of Milky Way satellites were uncorrelated then
Sagittarius would be expected to have a nearest neighbour as close or
closer than observed in $99$ out of $100$ random realizations
($P=0.99$). Sagittarius is thus relatively isolated and unlikely to be
a member of a physical pair.

\begin{figure*}
\centering
\hspace{-1.cm}
\resizebox{18cm}{!}
{\includegraphics{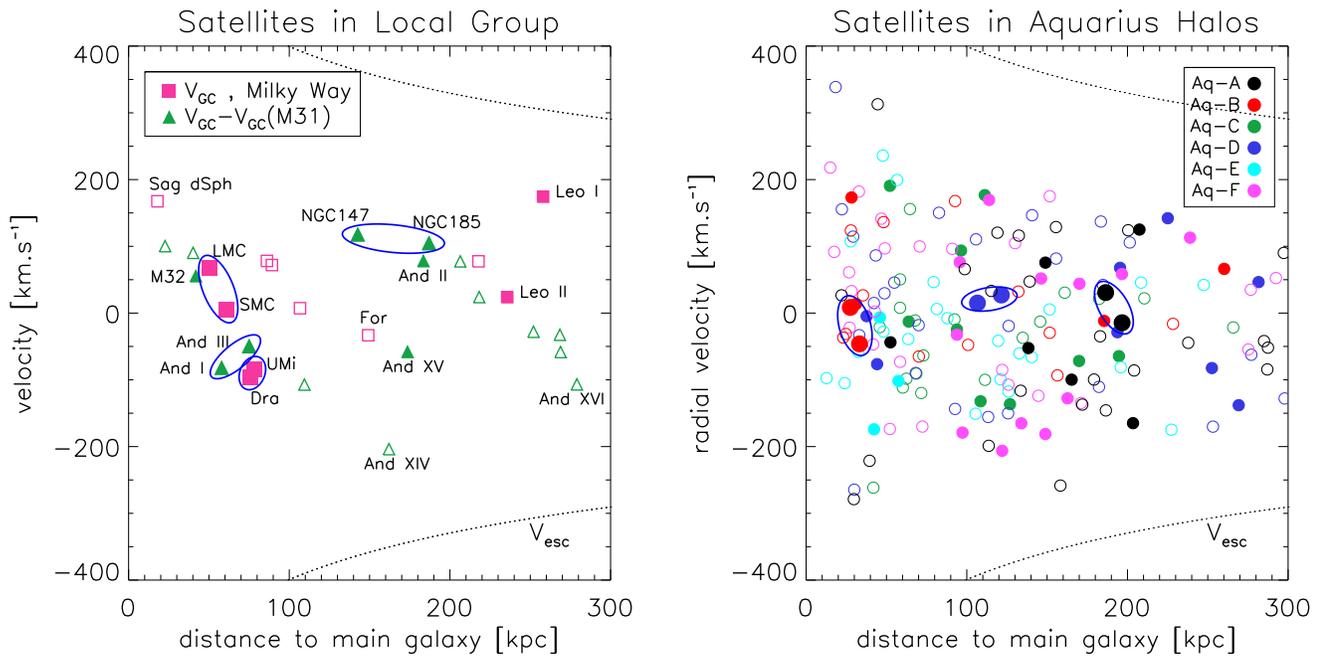}%
}\\%
\caption{{\it Left panel:} Galactocentric velocity versus distance for
  Local Group satellites brighter than $M_V=-8$. Distances are
  measured from the center of each primary: MW satellites are shown as
  magenta squares; M31's as green triangles. Velocities for the former
  are Galactocentric radial velocities; for the latter they refer to
  line-of-sight velocities relative to the 
  systemic velocity of M31. Dotted curves indicate, for reference, 
  the escape velocity
  from an NFW halo with virial velocity $V_{200}=250$ km/s and
  concentration $c=10$ \citep{Navarro1997}. Filled symbols highlight 
  satellites with a nearest neighbour
  much closer than expected by chance ($P<0.2$, see
  Fig.~\ref{FigP}). Pairs satisfying additional proximity criteria in
  velocity ($\Delta V<75$ km/s) and magnitude ($\Delta M_V<3$) are
  joined together by ellipses to indicate that they are likely
  physical pairs. These constitute $28\%$ of the total and include (i)
  the Magellanic Clouds; (ii) NGC 147 and NGC 185; (iii) Ursa Minor
  and Draco; and (iv) And I and And III. {\it Right panel:} Same as
  left panel, but for the semi-analytic satellite population of the six
  Aquarius halos. Different colors correspond to different
  halos. Note that the same
  criteria that pair $28\%$ of Local Group satellites link only six
  satellites in the Aquarius simulations, or just $3\%$ of the
  total.}
\label{FigfVR}
\end{figure*}

The situation reverses for Draco: its nearest neighbour, Ursa Minor,
lies only $23$ kpc away. This is much closer than expected at random;
a nearest neighbour that close occurs in fewer than $4$ out of $100$
random trials ($P=0.04$).  The right-hand panels of Fig.~\ref{FigDnn}
show as well two analogous examples for the M31 satellite
population. In this case, And IX is unlikely to be a member of a pair
($P=0.90$), whereas And III is unusually close to And I ($P=0.10$),
hinting at a possible physical association.

The distribution of the probability, $P$, that $d_{\rm nn}$ is as
small or smaller than observed if the radial and angular distribution
of Milky Way satellites were uncorrelated is shown by the solid and
dotted lines in Fig.~\ref{FigP} for the Milky Way and M31 satellites,
respectively.  Both curves are rather similar, indicating that the $P$
distribution is insensitive to the total number of satellites or their
radial distribution. It is also insensitive to assuming that the
satellite distribution is isotropic. Indeed, the solid and dotted curves in
Fig.~\ref{FigP} change almost imperceptibly if we confine the
Monte Carlo samples to a three-dimensional structure as flat as
observed for the Milky Way, i.e., roughly $3$:$1$ in its
major-to-minor axis ratio.

On the other hand, these distributions differ markedly from the
results for the Milky Way (shaded histogram in Fig.~\ref{FigP}) or the
combined M31+MW satellites (labelled ``Local Group'' in
Fig.~\ref{FigP}). A K-S test yields a probability of less than $0.3\%$
that the Local Group $P$ distribution is statistically consistent with
that of the random samples. There is a clear excess of
smaller-than-expected nearest-neighbour distances in the Local Group
that is difficult to account just by chance. For example, $45\%$ of
Local Group satellites have $P<0.2$ compared with the $20\%$ expected
if the distribution was isotropic.  

Our pairing procedure therefore retains all $P<0.2$ pairs (listed
horizontally in the labels of Fig.~\ref{FigP}) for further scrutiny.
A true physical pair must also differ little in velocity, so we impose
a maximum difference of $75$ km/s in the line-of-sight velocity
difference of the likely members. This threshold is motivated by the
velocity difference of the Magellanic Clouds, where there is little
doubt about their physical association. We assume for simplicity that
the same threshold applies regardless of the luminosity of the pair;
this is also consistent with the idea that most dwarfs should inhabit
halos of similar mass (see Sec.~\ref{SecIntro}) \citep[see][for a more
thorough discussion]{Sales2012}. Finally, since we are mainly
interested in pairs of comparable luminosities we retain as likely
pairs only those where their magnitudes differ by $\Delta M_{V}<3$.

Four pairs remain after applying these constraints, as shown in
Fig.~\ref{FigfVR}. Around the Milky Way,
aside from the Magellanic Clouds, the Ursa Minor and Draco pair is
singled out: they are $23$ kpc apart, and their velocities differ by
only $12$ km/s. Further, they have both approximately the same
luminosity, which makes them especially singular. The procedure also
joins together two pairs of satellites around M31: NGC 147 and NGC185,
as well as And I and And III. The latter are separated by $33$ kpc;
differ in velocity by just $32$ km/s; and in magnitude by $1.7$.

This analysis suggests that nearly $30\%$ of Local Group satellites
are in likely pairs ($8$ out of $29$). Of the four pairs, two
are almost indisputably associated (the Magellanic Clouds and
NGC 147/NGC 185; see, however, \citealt{Geha2010} for an alternate
view of the latter) but the other two might in principle result from
chance close encounters between unrelated satellites where projection
effects reduce the line-of-sight velocity difference. In order to
quantify these effects we have applied the same pairing procedure to
the satellite populations of the six Aquarius halos, as identified by
the semi-analytic model of \citet{Starkenburg2012}.

The blue dotted histogram in Fig.~\ref{FigP} shows that the $P$
distribution for Aquarius satellites differs little from that expected
from an isotropic distribution. Only $27\%$ of the $175$ satellites
have $P<0.2$; of those only $3$ pairs (i.e., fewer than $4\%$) pass as
well the velocity and magnitude criteria. The three Aquarius
  pairs (out of six halos) singled out by the analysis are shown in the right-hand panel
of Fig.~\ref{FigfVR}. Tracking their orbits back in time reveals that
none of them are actually physically related but that they result
simply from chance, transient associations in position and velocity
space. Note that this does not imply that {\it all} satellites
  have been accreted in isolation. As discussed by \citet{Wang2012b},
  a few of the bright satellites in Aquarius were accreted in groups,
  but the accretion happened early and the groups have long been disrupted
  by the tidal field of the main halo.  This
confirms the model expectation that physical associations amongst {\it
  satellites} should be extremely rare. The relatively high frequency
of satellite pairings in the Local Group indicates that the radial and
angular distributions of satellites are correlated, a fact that is not
easily accounted for by current dwarf galaxy formation models in the
$\Lambda$CDM paradigm.

\section{Summary and Conclusions}
\label{SecConc}

We have studied possible pairings amongst the satellites of the Milky
Way and of M31. Our procedure, which identifies unusually close
associations in position and velocity space, suggests that $8$ out of
the $29$ satellites brighter than $M_V=-8$ (i.e., nearly $\sim 30\%$)
form $4$ likely pairs of comparable luminosity ($\Delta
M_V<3$). These include the Magellanic Clouds; Ursa Minor and Draco;
NGC 147 and NGC 185; as well as And I and And III.

The same pairing procedure applied to a semi-analytic model of the
satellite population in the six halos of the Aquarius Project yields a 
likely pair fraction of fewer than $4\%$, even though the
model satellites have luminosity and radial distributions that match
closely that of the Local Group spirals. As expected, none of the
  Aquarius pairs correspond to true binary systems; rather, they
  result from transient associations between otherwise unrelated
satellites. The high pair frequency of the Local Group is unlikely to
be just a statistical fluke: the likely pair fraction of Aquarius
satellites never exceeds $12\%$ in thousands of random trials where
their magnitudes, angular directions and velocities are reshuffled.

We interpret these results as indicative of significant clustering in
the dwarf galaxy population of the Local Group. Although our analysis
only considers satellites brighter than $M_V = -8$ due to
incompleteness concerns and in order to allow comparison with
simulations, there have also been suggestions that some of the fainter
Galactic satellites are found in associations. In particular,
\citet{Belokurov2008} show that Leo IV and Leo V are close to each
other spatially and differ little in their line-of-sight
velocities. Further, the association of galaxies closest to the Local
Group, sometimes referred to as the loose NGC 3109 group
\citep{vandenBergh1999}, consists of four dwarf galaxies (NGC3109,
Antlia, Sextans A and B), two of which appear to be interacting
(NGC3019 and Antlia; see \citealt{Barnes2001}). These associations are
expected to be rare according to current $\Lambda$CDM galaxy formation
models, and may signal the presence of a mechanism that boosts the
likelihood of forming dwarf galaxies near other dwarfs, favouring some
environments over others. No such effect is present in current models
of dwarf galaxy formation, which rely largely on the mass accretion
history of individual halos to set the properties of a dwarf.

Candidate mechanisms that may bias the regions where dwarf galaxies
form in a way that enhances their clustering include patchy
reionization \citep[see, e.g.,][]{Lunnan2012}; large-scale
feedback effects on the surroundings of massive galaxies
\citep[][]{Alvarez2009,Busha2010a,Font2011a}; and the interaction of
dwarf galaxies with the cosmic web
\citep{Benitez-Llambay2012}. However, the effects of these models on
dwarf galaxies have yet to be developed fully and as a consequence the
importance of such effects on the dwarf galaxy population at large is
still unknown.

It is therefore important to firm up these findings (i) by extending
the analysis to fainter satellites, which should be possible once
photometric surveys of the northern and southern sky extend the
complete catalog of Milky Way satellites to fainter magnitudes;
(ii) by verifying, through accurate proper motion studies, that the
associations in position and line-of-sight velocities remain once the
full 6D phase space information is considered; (iii) by searching for
relic evidence of past interactions between likely pairs (such as the
Magellanic Stream for the LMC/SMC \citep[see,
e.g.,][]{Mathewson1974,Putman1998}; and, finally, (iv) by extending
this kind of analysis to a volume-limited survey of dwarf galaxy
associations in the local universe \citep[see, e.g.,][]{Karachentsev2008}. If confirmed, the enhanced
clustering of dwarfs may offer important clues to the formation of
faint galaxies that have yet to be identified and fully incorporated
into galaxy formation models.

\section*{Acknowledgments}

We are very grateful to Gabriella De Lucia, Amina Helmi and
Yang-Shyang Li for their role in developing the semi-analytic model of
galaxy formation used in this paper. We would like to thank the anonymous 
referee for comments which led to the improvement of the manuscript. 
ES is supported by the Canadian Institute 
for Advanced Research (CIfAR) Junior Academy and by a Canadian Institute for
Theoretical Astrophysics (CITA) National Fellowship.


\bibliographystyle{mn2e}
\bibliography{master}

\bsp

\label{lastpage}

\end{document}